\documentclass[intlimits,twoside,a4paper]{article}

\usepackage{amsmath,amssymb}
\usepackage{graphicx}
\usepackage{wrapfig}

\usepackage[T2A]{fontenc}
\usepackage[cp1251]{inputenc}
\usepackage{cmpj}

\issue{2011}{14}{4}{43702}

\doinumber{10.5488/CMP.14.43702}

\title[Non-perturbation theory of electronic dynamic conductivity]%
{Non-perturbation theory of electronic dynamic conductivity for
two-barrier resonance \\tunnel nano-structure}

\author[M.V.~Tkach, Ju.O.~Seti, O.M.~Voitsekhivska]{M.V.~Tkach\thanks{E-mail: ktf@chnu.edu.ua}\,, Ju.O.~Seti, O.M.~Voitsekhivska}
\address{Chernivtsi National University, 2 Kotsyubinsky Str.,
58012 Chernivtsi, Ukraine}

\date{Received August 29, 2011, in final form November 9, 2011}

\authorcopyright{M.V.~Tkach, Ju.O.~Seti, O.M.~Voitsekhivska, 2011}

\begin{document}

\maketitle

\begin{abstract}

The non-perturbation theory of electronic dynamic conductivity for
open two-barrier resonance tunnel structure is established for the
first time within the model of rectangular potentials and
different effective masses of electrons in the elements of
nano-structure and the wave function linear over the intensity of
electromagnetic field. It is proven that the results of the theory
of dynamic conductivity, developed earlier in weak signal
approximation within the perturbation method, qualitatively
and quantitatively correlate with the obtained results. The
advantage of non-perturbation theory is that it can be extended to
the case of electronic currents interacting with strong
electromagnetic fields in open multi-shell resonance tunnel
nano-structures, as active elements of quantum cascade lasers and
detectors.
\pacs 73.21.Fg, 73.90.+f, 72.30.+q, 73.63.Hs
\end{abstract}

\section{Introduction}

The experimental produce of nano-lasers and nano-detectors and,
further, quantum cascade lasers and detectors~\cite{1,2,3} stimulates
the intensive development of the theory of dynamic conductivity
for nano-heterosystems as active elements of these unique devices.
In spite of the twenty years period of investigating the
interaction between electromagnetic field and electronic currents
in open nano-structures, the respective theory is far from being
completed. One of the reasons is the mathematical problems
arising at the quantum mechanical research of physical processes
caused by the interaction of quasi-particles with classic and
quantized (phonons) fields in open nano-structures.

The theory of electronic conductivity for the two- and
three-barrier resonance tunnel structures (RTS)~\cite{4,5,6,7,8,9,10} is rather
complicated and mathematically sophisticated even without taking
into account the dissipative processes (scattering at the phonons,
impurities, imperfections). Therefore, the maximally   simplified
model is used in the above mentioned and other papers~\cite{11,12,13}:
$\delta$-like approximation of potential barriers for the
electrons and weak signal approximation equivalent to the first
order of perturbation theory (PT) over the intensity of
electromagnetic field interacting with electronic current in RTS.

We must note that $\delta$-like approximation of potential
barriers essentially simplifies the model of nano-structure.
Herein, the electron is automatically characterized by the unitary
effective mass within the whole system, which permits to calculate the
RTS dynamic conductivity~\cite{4,5,6,7,8,9,10,11,12,13} using the PT iteration method and
in such a way, quit the frames of linear approximation over the
field intensity.

Further, in references~\cite{14,15} it was shown that $\delta$-like
approximation of potential barriers correctly described the
qualitative properties of spectral parameters of quasi-stationary states of electrons
and the dynamic conductivity of
nano-structures but the magnitudes of resonance energies were
overestimated by tens per cent and resonance widths  by ten times with
respect to their magnitudes in a more realistic model of
rectangular potentials and different effective masses of electron
in the elements of nano-structure. It was displayed that in any
RTS, the $\delta$-barrier model strongly underestimated the
electrons life times in all quasi-stationary states and, thus, the
magnitude of the dynamic conductivity became the orders smaller
even for the structures with weak electromagnetic field.

The theory of dynamic conductivity established in references~\cite{16,17,18,19}
for the open two- and three-barrier RTS is based, as a rule, on a
more realistic model but it is so complicated compared to
the $\delta$-barrier model that it is practically
impossible to leave the framework of weak signal approximation.
Nevertheless, the development of experimental capabilities makes the
problem of strong interaction of electronic currents and
electromagnetic field in RTS more and more urgent.

Therefore, it is necessary to develop the non-perturbation theory
(NPT) of conductivity for open RTS where the intensity of
electromagnetic field would not play such a critical role as in PT.
The motivation for the positive expectations regarding the NPT
existed because the solution of complete Schrodinger equation with
Hamiltonian describing the interaction between electrons and
varying in time electromagnetic field was known [20]. However, in
spite of the known analytical expression for the exact wave
function, the theory of conductivity for the open RTSs was not
successfully developed.

The possible approach to the solution of this problem for the dynamic
conductivity of the two-barrier RTS is proposed in our paper for
the first time. We develop the NPT for the electronic dynamic
conductivity using the wave function which is the exact solution
of complete Schrodinger equation in the linear approximation over
the field. Being convinced  that the results obtained in the first order of PT
for electronic conductivity correlate with the results
of the herein developed NPT, the established approach can be used
in developing a general theory of electronic current interacting with
strong electromagnetic field in open multi-shell nano-structures,
being the active elements of quantum cascade lasers and detectors.

\section{Hamiltonian of the system. Finding the wave function from the complete Schrodinger equation}

\begin{figure}[!t]
\centerline{\includegraphics[width=0.45\textwidth]{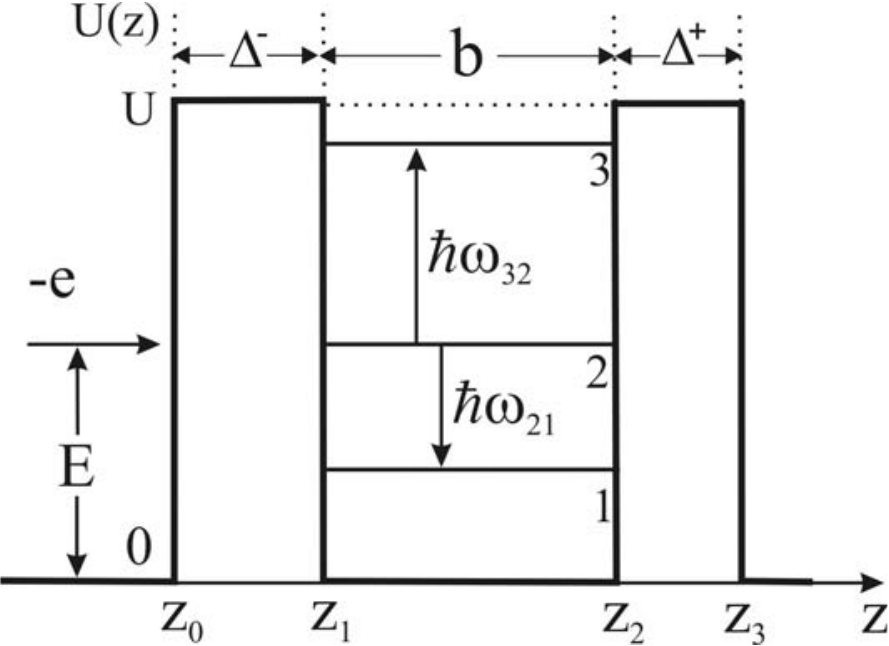}}
\caption{Energy scheme for the electrons and geometry of two-barrier RTS.} \label{fig1}
\end{figure}

The open two-barrier RTS is studied in the Cartesian coordinate
system with OZ axis perpendicular to the planes of nano-structure,
figure~\ref{fig1}.
The small difference between the lattice constants of
the nano-structure wells and barriers makes it possible to use the effective
masses
\begin{equation}
\label{eq1} m(z)=m_{0} \sum\limits_{p = 0}^{2} [\theta
(z-z_{2p-1})-\theta (z-z_{2p})]+m_{1} \sum\limits_{p = 0}^{1}
[\theta (z-z_{2p})-\theta (z-z_{2p+1})]
\end{equation}
and rectangular potentials
\begin{equation}
\label{eq2} U(z)=U \sum\limits_{p = 0}^{1} [\theta
(z-z_{2p})-\theta (z-z_{2p+1})]  .
\end{equation}
Here $\theta (z)$ is Heaviside step function; $z_{-1}\rightarrow -
\infty $, $z_{4}\rightarrow + \infty$.

It is assumed that mono-energetic electron current with the
energy ($E$), density of current ($J_{0}^{+} \sim \sqrt{E}$) and
concentration ($n_{0}$) moving perpendicularly to the planes of
two-barrier RTS falls at it from the left side. The electronic
movement can be considered as one-dimensional
($\overrightarrow{k}_{||}=0$). According to the numeric evaluations,
the velocity within the nano-structure is by 3--4 orders smaller
than the velocity outside. Thus, the interaction between electrons
and electromagnetic field with frequency ($\omega$) and intensity
of electric field ($\epsilon$) is essential only within the two-barrier
RTS and can be neglected outside it.

The electron wave function has to satisfy the complete Schrodinger
equation
\begin{equation}
\label{eq3} \mathrm{i} \hbar \frac{\partial \Psi (E, \omega, z,
t)}{\partial t}= [H_{0}(z)+H_{1}(z, t)] \Psi (E, \omega, z, t)  ,
\end{equation}
where
\begin{equation}
\label{eq4} H_{0}(z)=-\frac{\hbar^{2}}{2} \frac{\partial}{\partial
z} \frac{1}{m(z)} \frac{\partial}{\partial z} + U(z)
\end{equation}
is the Hamiltonian of the electron without interaction with the field.

The electron interaction with the electromagnetic
field varying in time, is described by the Hamiltonian
\begin{equation}
\label{eq5} H_{1}(z, t)=-2 e \epsilon z [\theta (z)-\theta
(z-z_{3})] \cos \omega t .
\end{equation}

The both linearly independent exact solutions of complete
Schrodinger equation with Hamiltonian $H_{0}(z)$ in the potential
well are known: $\exp (\pm \mathrm{i} k z - \mathrm{i} \omega_{0}
t)$ [17, 19], $\omega_{0} = E \hbar^{-1}$. The both linearly
independent exact solutions of equation (\ref{eq3}) with Hamiltonian
$H(z,t)=H_{0}(z)+H_{1}(z,t)$, taking into account the
electron-electromagnetic field interaction in linear approximation
over the electric intensity are also known:
\[\exp \left[
\mathrm{i} \left(\pm k z - \omega_{0} t + \frac{2 e \epsilon z}{\hbar
\omega} \sin \omega t \pm \frac{2 e \epsilon k}{m \omega^{2}} \cos
\omega t\right)\right],
\]
where $k$ is electron quasi-momentum~\cite{20}. Thus,
using the exact solutions of equation (\ref{eq3}) for the case of
two-barrier RTS, the electron wave function is written as
\begin{equation}
\label{eq6} \Psi (E, \omega, z, t)=\Psi_{0} (E, z, t) \theta
(-z) +
\sum\limits_{p = 1}^{3} \Psi_{p} (E, \omega, z, t) [\theta
(z-z_{p-1})-\theta (z-z_{p})] + \Psi_{4} (E, z, t) \theta
(z-z_{3}).
\end{equation}

In the outer media of two-barrier RTS, where the interaction with
electromagnetic field is neglected, the wave functions are
\begin{equation}
 \label{eq7}
 \Psi_{0} (E, z, t) = (a_{0} \mathrm{e}^{\mathrm{i} k z} + b_{0}
\mathrm{e}^{- \mathrm{i} k z}) \mathrm{e}^{- \mathrm{i} \omega_{0}
t}  ,
\end{equation}
\begin{equation}
 \label{eq8}
  \Psi_{4} (E, z, t) = a_{4} \mathrm{e}^{\mathrm{i} k z-\mathrm{i}
\omega_{0} t}  ,
\end{equation}
where  \[k=\hbar^{-1} \sqrt{2 m_{0} E} \, . \]

Here it is taken into account that the mono-energetic electronic
current impinges at RTS from the left hand side, since in the left hand media
there is both a falling and a reflected wave while in the right one there is
the wave moving towards the infinity only.

Within the two-barrier RTS, where the electron-electromagnetic
field interaction is essential, the wave function is found as
linear combinations of eigen wave functions of the Hamiltonian
$H(z,t)$
\begin{align}
 \label{eq9}
 \Psi_{1} (E, \omega, z, t)& = \left[a_{1} f_{1}^{\rightarrow} (E, \omega, z, t) {\re}^{\chi z} + b_{1} f_{1}^{\leftarrow} (E, \omega, z, t) \mathrm{e}^{- \chi
z}\right] \mathrm{e}^{- \mathrm{i} \omega_{0} t}  ,\\
 \label{eq10}
 \Psi_{2} (E, \omega, z, t) &= \left[a_{2} f_{2}^{\rightarrow} (E, \omega, z, t) \mathrm{e}^{\mathrm{i} k z} + b_{2} f_{2}^{\leftarrow} (E, \omega, z, t) \mathrm{e}^{-
\mathrm{i} k z}\right] \mathrm{e}^{- \mathrm{i} \omega_{0} t} ,\\
 \label{eq11}
 \Psi_{3} (E, \omega, z, t) &= \left[a_{3} f_{1}^{\rightarrow} (E,
\omega, z, t) \mathrm{e}^{\chi z} + b_{3} f_{1}^{\leftarrow} (E,
\omega, z, t) \mathrm{e}^{- \chi z}\right] \mathrm{e}^{- \mathrm{i}
\omega_{0} t}  ,
\end{align}
where
\begin{equation}
\label{eq12} f_{p}^{\rightleftarrows} (E,
\omega, z, t) = \exp (\mathrm{i} \alpha \sin \omega t \pm
\mathrm{i} \beta_{p} \cos \omega t) \ , \quad \quad (p=1, 2)
\end{equation}
\begin{equation}
\label{eq13}
\displaystyle \alpha = \frac{2 e \epsilon z}{\hbar \omega}\,; \quad \quad \beta_{1} =
\frac{2 \mathrm{i} e \epsilon \chi}{m_{1} \omega^{2}}\,; \quad \quad
\beta_{2} = \frac{2 e \epsilon k}{m_{0} \omega^{2}}\,; \quad \quad
\chi=\hbar^{-1} \sqrt{2 m_{1} (U-E)} \, .
\end{equation}

Further, using the known~\cite{21} expansion of exponential functions into
Fourier range over all the harmonics, the functions
$f_{p}^{\rightleftarrows} (E, \omega, z, t)$ are written as
\begin{equation}
\label{eq14} f_{p}^{\rightleftarrows} (E, \omega, z, t) =
\sum\limits_{n_{1} = - \infty}^{\infty} \sum\limits_{n_{2} = -
\infty}^{\infty} \mathrm{i}^{\pm n_{2}} j_{n_{1}} (\alpha)
j_{n_{2}} (\beta_{p}) \mathrm{e}^{\mathrm{i} (n_{1} \mp n_{2})
\omega t} \ ,
\end{equation}
where $j_{n}$ are the cylindrical Bessel functions of the whole
order.

The quantum transitions, accompanied by energy radiation or
absorption, occur between electron quasi-stationary states with
odd number under the effect of electromagnetic field. The most
intensive transitions arise between the neighbouring resonance
states. Thus, the expression (\ref{eq14}) for the functions
$f_{p}^{\rightleftarrows} (E, \omega, z, t)$ can be essentially
simplified by leaving only zero and first harmonics from the whole
infinite range. Then, in a one-mode approximation for the functions
$f_{p}^{\rightleftarrows} (E, \omega, z, t)$, the following
expression is obtained
\begin{eqnarray}
\label{eq15} f_{p}^{\rightleftarrows} (E, \omega, z, t) &=&
C_{p}^{\rightleftarrows} (E, \omega, z) + \left\{D_{p}(E, \omega, z) +
\mathrm{i} \left[F_{p}^{\rightleftarrows} (E, \omega, z) +
G_{p}^{\rightleftarrows} (E, \omega, z)\right]\right\} \mathrm{e}^{- \mathrm{i}
\omega t}
\nonumber\\
&+& \left\{-D_{p}(E, \omega, z) + \mathrm{i} \left[F_{p}^{\rightleftarrows} (E,
\omega, z) - G_{p}^{\rightleftarrows} (E, \omega, z)\right]\right\}
\mathrm{e}^{\mathrm{i} \omega t}  ,
\end{eqnarray}
with
\begin{equation}
\label{eq16} C_{p}^{\rightleftarrows} (E, \omega, z) =
j_{0}(\alpha) j_{0}(\beta_{p}) + 2 \sum\limits_{n_{1} =
1}^{\infty} [j_{4 n_{1}}(\alpha) j_{4 n_{1}}(\beta_{p})-j_{4
n_{1}-2}(\alpha) j_{4 n_{1}-2}(\beta_{p})]  ,
\end{equation}
\begin{equation}
\label{eq17}
D_{p}(E, \omega, z)=\sum\limits_{n_{1} = 1}^{\infty} \{j_{4
n_{1}-1}(\alpha) [j_{4 n_{1}}(\beta_{p})+j_{4 n_{1}-2}(\beta_{p})]
- j_{4 n_{1}-3}(\alpha) [j_{4 n_{1}-2}(\beta_{p})+j_{4
n_{1}-4}(\beta_{p})]\}  ,
\end{equation}
\begin{equation}
\label{eq18}
G_{p}^{\rightleftarrows} (E, \omega, z) = \pm \sum\limits_{n_{1}
= 1}^{\infty} \{j_{4 n_{1}}(\alpha) [j_{4 n_{1}-1}(\beta_{p})-j_{4
n_{1}+1}(\beta_{p})]+j_{4 n_{1}-2}(\alpha) [j_{4
n_{1}-1}(\beta_{p})-j_{4 n_{1}-3}(\beta_{p})]\} .
\end{equation}

Within the framework of the linear Hamiltonian over the field intensity
($\epsilon$), the rather complicated formulas (\ref{eq16})--(\ref{eq18}) correctly define
the electron wave function in a one-mode approximation independently
of the intensity magnitude. In the case of small intensity when the
condition
\begin{equation}
\label{eq19} \min [\alpha(E), \beta_{1}(E), \beta_{2}(E)]\ll 1
\end{equation}
is fulfilled, expanding the Bessel functions into a series and
preserving the linear  term over the field, the expressions for
coefficients are simplified
\begin{equation}
\label{eq20} C_{p}^{\rightleftarrows}=1; \quad \quad \quad D_{p}=-\alpha /2;
\quad \quad \quad F_{p}^{\rightleftarrows}=\pm \beta_{p}/2; \quad \quad \quad
G_{p}^{\rightleftarrows}=0  .
\end{equation}

Thus, the wave function is also obtained in a convenient analytical
form
\begin{eqnarray}
 \Psi(E, \omega, z, t)\! &=& \!(a_{0} \mathrm{e}^{\mathrm{i}
k z}+b_{0} \mathrm{e}^{- \mathrm{i} k z}) \mathrm{e}^{- \mathrm{i}
\omega_{0} t} \theta(-z) + a_{4} \mathrm{e}^{\mathrm{i} k z}
\mathrm{e}^{- \mathrm{i} \omega_{0} t} \theta(z-z_{3})
\nonumber\\
&+&\! \sum\limits_{p = 1}^{3} \mathrm{e}^{- \mathrm{i} \omega_{0} t}
\left\{ a_{p} \mathrm{e}^{K_{p} z}\left[1+\frac{1}{2} (\mathrm{i}
\beta_{p}+\alpha) \mathrm{e}^{\mathrm{i} \omega t} + \frac{1}{2}
(\mathrm{i} \beta_{p}-\alpha) \mathrm{e}^{-\mathrm{i} \omega t}\right]\right.
\nonumber\\
&+&\! \left.b_{p} \mathrm{e}^{- K_{p} z}[1-\frac{1}{2} (\mathrm{i}
\beta_{p}-\alpha) \mathrm{e}^{\mathrm{i} \omega t} - \frac{1}{2}
(\mathrm{i} \beta_{p}+\alpha) \mathrm{e}^{-\mathrm{i} \omega t}]\right\}
\left[\theta (z - z_{p-1})-\theta (z - z_{p})\right] , \quad \label{eq21}
\end{eqnarray}
where \[K_{1} = K_{3} = \chi; \quad \quad \quad K_{2} =\mathrm{i} k  . \]

Using the obtained wave function $\Psi(E, \omega, z, t)$ one can
perform the calculation of the permeability coefficient for the
two-barrier RTS, obtaining the spectral parameters of electron
quasi-stationary states and active dynamic conductivity of
nano-structure.

\section{Permeability coefficient and dynamic conductivity of two-barrier RTS}

Now, we can find the dynamic conductivity caused by the
quantum transitions of electrons from the quasi-stationary state with the
energy $E$ into the states with the energies $E+\hbar \omega$ or
$E-\hbar \omega$ due to the effect of the periodical
electromagnetic field with intensity $\epsilon$ and frequency
$\omega$. Therefore, we have to define the densities of electron
currents: $J(E+\hbar \omega)$ and $J(E-\hbar \omega)$, flowing out
of the RTS with the respective energies. In such approach, the
complete wave function is written as linear combination of wave
functions describing the electron states with the energies $E$,
$E+\hbar \omega$ and $E-\hbar \omega$. The functions $\Psi(E \pm
\hbar \omega, \omega, z, t)$ can be obtained from the expression
(\ref{eq21}) for the already known function $\Psi(E, \omega, z, t)$ using
the substitution $E \rightarrow E \pm \hbar \omega$. Then,
\begin{eqnarray}
\label{eq22} \Psi(E \pm \hbar \omega, \omega, z, t) &=& b_{0}^{\pm}
\mathrm{e}^{- \mathrm{i} k^{\pm} z} \mathrm{e}^{- \mathrm{i}
(\omega_{0} \pm \omega) t} \theta(-z) + a_{4}^{\pm}
\mathrm{e}^{\mathrm{i} k^{\pm} z} \mathrm{e}^{- \mathrm{i}
(\omega_{0} \pm \omega) t} \theta(z-z_{3})
\nonumber\\
&+& \sum\limits_{p = 1}^{3} \mathrm{e}^{- \mathrm{i} \omega_{0} t}
\left\{ a_{p}^{\pm} \mathrm{e}^{K_{p}^{\pm} z}\left[\mathrm{e}^{\mp
\mathrm{i} \omega t}+\frac{1}{2} (\mathrm{i} \beta_{p}^{\pm} \pm
\alpha) + \frac{1}{2} (\mathrm{i} \beta_{p}^{\pm} \mp \alpha)
\mathrm{e}^{\mp 2 \mathrm{i} \omega t}\right]\right.
\nonumber\\
&+& \left.b_{p}^{\pm} \mathrm{e}^{- K_{p}^{\pm} z}\left[\mathrm{e}^{\mp
\mathrm{i} \omega t}-\frac{1}{2} (\mathrm{i} \beta_{p}^{\pm} \mp
\alpha) - \frac{1}{2} (\mathrm{i} \beta_{p}^{\pm} \pm \alpha)
\mathrm{e}^{\mp 2 \mathrm{i} \omega t}\right]\right\}
\nonumber\\
&\times&\left[\theta (z -
z_{p-1})+\theta (z - z_{p})\right] ,
\end{eqnarray}
where
\begin{equation}
\label{eq23} K_{1}^{\pm} = K_{3}^{\pm} = \chi^{\pm} = \hbar^{-1}
\sqrt{2 m_{1} [U-(E \pm \hbar \omega)]} \, ; \quad K_{2}^{\pm} =
\mathrm{i} k^{\pm} = \mathrm{i} \hbar^{-1} \sqrt{2 m_{0} (E \pm
\hbar \omega)} \, ;
\end{equation}
\begin{equation}
\label{eq24}
\displaystyle \beta_{1}^{\pm} = \frac{2 \mathrm{i} e \epsilon \chi^{\pm}}{m_{1}
\omega^{2}}\,; \qquad  \beta_{2}^{\pm} = \frac{2 e \epsilon
k^{\pm}}{m_{0} \omega^{2}} \, .
\end{equation}

The mono-energetic electron current falls at RTS with the energy
$E=\hbar \omega_{0}$. Under the effect of electromagnetic field
there occur quantum transitions into higher (with the energy
$E+\hbar \omega$) or lower (with the energy $E-\hbar \omega$)
electron quasi-stationary states, the currents from which produce
the dynamic conductivity of a nano-structure.  In order to describe
this physical process correctly, we have to leave the terms
containing only the first harmonic ($\pm \omega$) in formula (\ref{eq22})
for the wave function. Thus,
\begin{eqnarray}
\label{eq25} \Psi(E \pm \hbar \omega, \omega, z, t) &=& b_{0}^{\pm}
\mathrm{e}^{- \mathrm{i} k^{\pm} z} \mathrm{e}^{- \mathrm{i}
(\omega_{0} \pm \omega) t} \theta(-z) + a_{4}^{\pm}
\mathrm{e}^{\mathrm{i} k^{\pm} z} \mathrm{e}^{- \mathrm{i}
(\omega_{0} \pm \omega) t} \theta(z-z_{3})
\nonumber\\
&+& \sum\limits_{p = 1}^{3} \mathrm{e}^{- \mathrm{i} (\omega_{0} \pm
\omega) t} ( a_{p}^{\pm} \mathrm{e}^{K_{p}^{\pm} z} + b_{p}^{\pm}
\mathrm{e}^{- K_{p}^{\pm} z})[\theta (z - z_{p-1})+\theta (z -
z_{p})]  .
\end{eqnarray}

The complete wave function $\Phi (E, E-\hbar \omega, E+\hbar
\omega,\omega, z, t)$ can be written at the base of superposition
(linear combination) of wave functions (\ref{eq21}) and (\ref{eq25}). It depends
on electron energies $E$, $E-\hbar \omega$, $E+\hbar \omega$ and
electromagnetic field frequency $\omega$. For a convenient
presentation, it is further written as $\Phi (E, \omega, z, t)$.
\begin{eqnarray}
\label{eq26} \Phi (E, \omega, z, t)&=&\Phi_{0} (E, \omega, z, t) \theta
(-z) \nonumber\\
&+&
\sum\limits_{p = 1}^{3} \Phi_{p} (E, \omega, z, t) [\theta
(z-z_{p-1})-\theta (z-z_{p})] + \Phi_{4} (E, \omega, z, t) \theta
(z-z_{3}) ,
\end{eqnarray}
where
\begin{align}
\label{eq27} \Phi_{0}(E, \omega, z, t) &= \mathrm{e}^{- \mathrm{i}
\omega_{0} t} \left(A_{0} \mathrm{e}^{\mathrm{i} k z} + B_{0}
\mathrm{e}^{- \mathrm{i} k z} + B_{0}^{+} \mathrm{e}^{- \mathrm{i}
k^{+} z} \mathrm{e}^{- \mathrm{i} \omega t} + B_{0}^{-}
\mathrm{e}^{- \mathrm{i} k^{-} z} \mathrm{e}^{\mathrm{i} \omega
t}\right) ,\\
\label{eq28}
\Phi_{p}(E, \omega, z, t) &= \mathrm{e}^{- \mathrm{i} \omega_{0} t}
\left\{ A_{p} \mathrm{e}^{K_{p} z}\left[1+\frac{1}{2} (\mathrm{i}
\beta_{p}+\alpha) \mathrm{e}^{\mathrm{i} \omega t} + \frac{1}{2}
(\mathrm{i} \beta_{p}-\alpha) \mathrm{e}^{-\mathrm{i} \omega t}\right]\right.
\nonumber\\
&+ B_{p} \mathrm{e}^{- K_{p} z}\left[1-\frac{1}{2} (\mathrm{i}
\beta_{p}-\alpha) \mathrm{e}^{\mathrm{i} \omega t} - \frac{1}{2}
(\mathrm{i} \beta_{p}+\alpha) \mathrm{e}^{-\mathrm{i} \omega t}\right]
\nonumber\\
&+ \left.\left( A_{p}^{+} \mathrm{e}^{K_{p}^{+} z} + B_{p}^{+} \mathrm{e}^{-
K_{p}^{+} z}\right) \mathrm{e}^{-\mathrm{i} \omega t} + \left( A_{p}^{-}
\mathrm{e}^{K_{p}^{-} z} + B_{p}^{-} \mathrm{e}^{- K_{p}^{-} z}\right)
\mathrm{e}^{\mathrm{i} \omega t} \right\} ,\\
\label{eq29}
\Phi_{4}(E, \omega, z, t)& = \mathrm{e}^{- \mathrm{i} \omega_{0} t}
\left(A_{4} \mathrm{e}^{\mathrm{i} k z} + A_{4}^{+}
\mathrm{e}^{\mathrm{i} k^{+} z} \mathrm{e}^{- \mathrm{i} \omega t}
+ A_{4}^{-} \mathrm{e}^{\mathrm{i} k^{-} z} \mathrm{e}^{
\mathrm{i} \omega t}\right)  .
\end{align}

The two-barrier RTS under study is an open one,
consequently the wave function $\Phi (E, \omega, z, t)$ at any moment
of time has to satisfy the normality condition
\begin{equation}
\label{eq30} \int\limits_{ - \infty} ^{\infty}  \Phi^{*} (k',
\omega, z, t) \Phi(k, \omega, z, t) \mathrm{d} z = \delta (k - k')  .
\end{equation}
The wave function and its density of current should be continuous
at all nano-structure interfaces
\begin{equation}
\label{eq31} \displaystyle \Phi_{p}(E, \omega, z_{p}, t) = \Phi_{p+1}(E, \omega,
z_{p}, t); \quad \quad \frac{\partial \Phi_{p}(E, \omega, z, t)}{m_{0
(1)} \
\partial z}\bigg|_{z=z_{p}} = \frac{\partial \Phi_{p+1}(E, \omega, z, t)}{m_{1 (0)} \
\partial z}\bigg|_{z=z_{p}}  .
\end{equation}

The coefficients at zero harmonics: $B_{0}, A_{p}, B_{p}, A_{4}$
are definitely obtained from the system of homogeneous equations
(\ref{eq31}) through the coefficient $A_{0}$. This is, in turn, related
to the density of start electron current impinging at RTS:
$J_{0}^{+}=e n_{0} \sqrt{2 E m_{0}^{-1}} |A_{0}|^{2}$, where
$n_{0}$ is the concentration of electrons in this current, $e$ is
electron charge. Coefficients at the first harmonics:
$B_{0}^{\pm}, A_{p}^{\pm}, B_{p}^{\pm}, A_{4}^{\pm}$ are defined
through the now known coefficients at zero harmonics of function
$\Phi (E, \omega, z, t)$. According to the quantum mechanics~\cite{24},
the permeability coefficient for the two-barrier RTS is
\begin{equation}
\label{eq32} D(E)=|A_{4}/A_{0}|^{2}  .
\end{equation}

It is well known~\cite{14,22} that the permeability coefficient $D(E)$
 determines the spectral parameters: resonance energies ($E_{n}$) and
 resonance widths ($\Gamma_{n}$) of  quasi-stationary states of electrons.
The positions of $D(E)$ maxima in the energy scale fix the resonance
 energies, while their widths at the halves of maximal heights $D(E_{n})$
 fix the resonance widths of these quasi-stationary states.

 According to electrodynamics~\cite{23}, in a quasi-static approximation,
 the energy ($\mathcal{E}$), got by the electrons from the field during
 the period $T=2 \pi / \omega$, is related to the real part of dynamic
 conductivity ($\sigma$)
\begin{equation}
\label{eq33} \mathcal{E} = \frac{4 \pi z_{3} \epsilon^{2}}{\omega}
\sigma (E, \omega)  .
\end{equation}

The same energy is defined by the electron currents flowing out
of the nano-structure through the densities of currents of
uncoupling electrons
\begin{equation}
\label{eq34} \mathcal{E} = \frac{\hbar \omega T}{e} \left\{\left[J(E+\hbar
\omega, z_{3})-J(E+\hbar \omega, 0)\right]-\left[J(E-\hbar \omega,
z_{3})-J(E-\hbar \omega, 0)\right]\right\}  .
\end{equation}

According to quantum mechanics~\cite{24}, the density of current
is defined by the wave function $\Phi (E, \omega, z, t)$
\begin{equation}
\label{eq35} J(E, z) = \frac{\mathrm{i} e \hbar n_{0}}{2 m(z)}
\left[\Phi(E, \omega, z, t) \frac{\partial \Phi^{*}(E, \omega, z, t)
}{\partial z}-\Phi^{*}(E, \omega, z, t) \frac{\partial \Phi(E,
\omega, z, t) }{\partial z}\right]  .
\end{equation}
Thus, as a result of analytical calculations, the final expression
for the dynamic conductivity of two-barrier RTS is obtained:
\begin{equation}
\label{eq36} \sigma(E, \omega) = \frac{\hbar^{2} \omega n_{0}}{2
z_{3} m_{0} \epsilon^{2}} [k^{+} (|B_{0}^{+}|^{2} +
|A_{4}^{+}|^{2}) - k^{-} (|B_{0}^{-}|^{2} + |A_{4}^{-}|^{2})] \ .
\end{equation}

It is evident that in linear approximation over the electromagnetic
field intensity, the coefficients $B_{0}^{\pm}, A_{4}^{\pm}$ are also linear.
Consequently, in this approximation, (the same in PT~\cite{4,5,6,7,8,9,10,11,12,13,15,16,17,18,19})
the dynamic conductivity is independent of $\epsilon$.

The calculations of spectral parameters of electron
quasi-stationary states and dynamic \linebreak conductivity of nano-structure
were performed at the base of the developed NPT for \linebreak
In$_{0.52}$Al$_{0.48}$As/In$_{0.53}$Ga$_{0.47}$As two-barrier RTS
with physical parameters: $m_{0}$ = 0.046~$m_{e}$, $m_{1}$ =
0.089~$m_{e}$, $U$ = 516~meV, $n_{0}$ = $10^{16}$ cm$^{-3}$ and
typical geometrical ones: $b$ = 10.8~nm, $\Delta^{+}+\Delta^{-}$ =
6~nm, $\Delta^{+}$ = 2 $\div$ 4.5~nm. The same calculations were
performed in the frames of the previously developed PT~\cite{19} in the
first order over the field intensity  for comparison.

The results obtained for positive or negative conductivities
$\sigma (E, \Omega=\hbar \omega)$ produced by the quantum
transitions of electrons interacting with electromagnetic field
in the processes of absorption ($1 \rightarrow 2$) or radiation
($2 \rightarrow 1$) are shown in figures~2~(a),~(b) and figures~2~(d),~(e), respectively.

The spectral parameters (resonance energies and resonance widths)
of electron quasi-stationary states, defined from the permeability
coefficient~\cite{14,22}, do not depend on the method of  calculation.
Their magnitudes, obtained for the two-barrier RTS with $b$=10.8
nm, $\Delta^{+}=\Delta^{-}$=3~nm are presented in
figures~2~(a), (b), (d), (e). The figures prove that the functions $\sigma
(E, \Omega={\rm const})$ and $\sigma (E={\rm const}, \Omega)$ are of the shape of
Lorentz curve in both methods (NPT and PT). However, herein, it is
clear that the magnitudes $\sigma (E, \Omega)$ in PT are
overestimated in the detector and underestimated in laser quantum
transitions at any $E$ and $\omega$, comparing to the exacter NPT.

In the both methods, the positive conductivity maximum:
$\sigma_{12}=\max \sigma_{12} (E, \Omega)$, caused by the detector
(accompanied by electromagnetic wave absorption) quantum
transitions between the first and second quasi-stationary states
is calculated at the plane ($E$, $\Omega$) in the point:
$E=E_{1}$, $\Omega=\Omega_{12}=E_{2}-E_{1}$. For the two-barrier
RTS under study we obtained: $\sigma_{12}^{\rm PT}$=20557~S/cm, and
$\sigma_{12}^{\rm NPT}$=16740~S/cm. It means that the PT gives the
magnitude at $22.8~\% $ bigger than the NPT. It is also shown
that in the both methods, the widths ($\Gamma_{\Omega}$) of $\sigma
(E_{1}, \Omega)$ functions are almost coinciding and in $\Omega$
scale coincide to the resonance width of the second quasi-stationary
state ($\Gamma_{\Omega}^{\rm NPT} \approx \Gamma_{\Omega}^{\rm PT} \approx
\Gamma_{2}$). The widths ($\Gamma_{E}$) of $\sigma (E,
\Omega_{12})$ functions coincide in $E$ scale and with
resonance width of the first quasi-stationary state
($\Gamma_{E}^{\rm NPT} \approx \Gamma_{E}^{\rm PT} \approx \Gamma_{1}$).

In the both methods, the negative conductivity minimum:
$\sigma_{21}=\min \sigma_{21} (E, \Omega)$, caused by the laser
(accompanied by electromagnetic wave radiation) quantum transitions
between the second and first quasi-stationary states is placed at the
plane ($E$, $\Omega$) in the point: $E=E_{2}$, $\Omega=\Omega_{21}=E_{2}-E_{1}$.
For the two-barrier RTS under study, we obtained: $\sigma_{21}^{\rm PT}=-41300$~S/cm,
$\sigma_{21}^{\rm NPT}=-49990$~S/cm. Contrary to the positive conductivity,
the widths of the negative one in both scales ($E$, $\Omega$) are close
to each other and to the resonance width $\Gamma_{1}$, i.e.
$\Gamma_{E}^{\rm NPT} \approx \Gamma_{E}^{\rm PT} \approx \Gamma_{\Omega}^{\rm NPT}
\approx \Gamma_{\Omega}^{\rm PT} \approx \Gamma_{1}$.

In figures~2~(c),~(f) the dependences of maximal (minimal) magnitudes of positive
(negative) conductivities on the width of the outer barrier ($\Delta^{+}$)
at a fixed sum width of both barriers ($\Delta^{+}+\Delta^{-}=6$~nm)
are shown for NPT and PT. It is clear that the functions in the transitions
$1 \leftrightarrows 2$ and $2 \leftrightarrows 3$ are located close
to each other in both methods not only qualitatively but also quantitatively.
The insertions in figures~2~(c), (f) prove that the errors
($\eta = 1-\sigma^{\rm PT}/\sigma^{\rm NPT} , \% $) of conductivities calculated
within PT with respect to NPT weakly depend on the relationship between the
widths of both barriers.

Finally, we should note that in the both methods the dependences of
$\sigma$ on $\Delta^{+}$ at $\Delta^{-}+\Delta^{+}=\Delta={\rm const}$ are
not only of the same shape (figures~2~(c), (f)) but their magnitudes
at any $\Delta^{+}$ are close to each other. The behaviour of the
function is clearly explained by physical considerations. In
reference~\cite{19} it was proven that the magnitude of dynamic
conductivity is proportional to the electron life times in those
quasi-stationary states between which the quantum transitions
occur due to the interaction with electromagnetic field. If the
difference between the widths of both barriers
$|\Delta^{+}-\Delta^{-}|$ is big, the life times in all
quasi-stationary states are small because the electrons rapidly
quit the two-barrier RTS through the thinner barrier. If the
barriers widths correlate, the life times in all
quasi-stationary states increase, approaching the maximal
magnitude at $\Delta^{-}=\Delta^{+}=\Delta/2$. From figures~2~(c), (f)
one can see that $\sigma$ dependence on $\Delta^{+}$ is
qualitatively similar to the above described evolution of life
times with the only difference that $\max \sigma (\Delta^{+})$ is
approached not at $\Delta^{-}=\Delta^{+}=\Delta/2$ but at
$\Delta_{0}^{+}>\Delta/2$. This is also clear, because, contrary
to the electron life times in quasi-stationary states
independent on the number of electrons in two-barrier RTS (in the
frames of the model neglecting electron-electron interaction), the
conductivity depends on this factor due to the interaction with
electromagnetic field. Thus, at the increase of input barrier
width ($\Delta^{-}$), the number of electrons reflected from RTS
increases. Therefore, at
$\Delta^{-}=\Delta+\Delta_{0}^{+}<\Delta/2$ the number of
electrons in the RTS is bigger than at
$\Delta^{-}=\Delta^{+}=\Delta/2$ and, consequently,
$\sigma(\Delta_{0}^{+})>\sigma(\Delta/2)$.

\begin{figure}[!h]
\centerline{\includegraphics[width=0.9\textwidth]{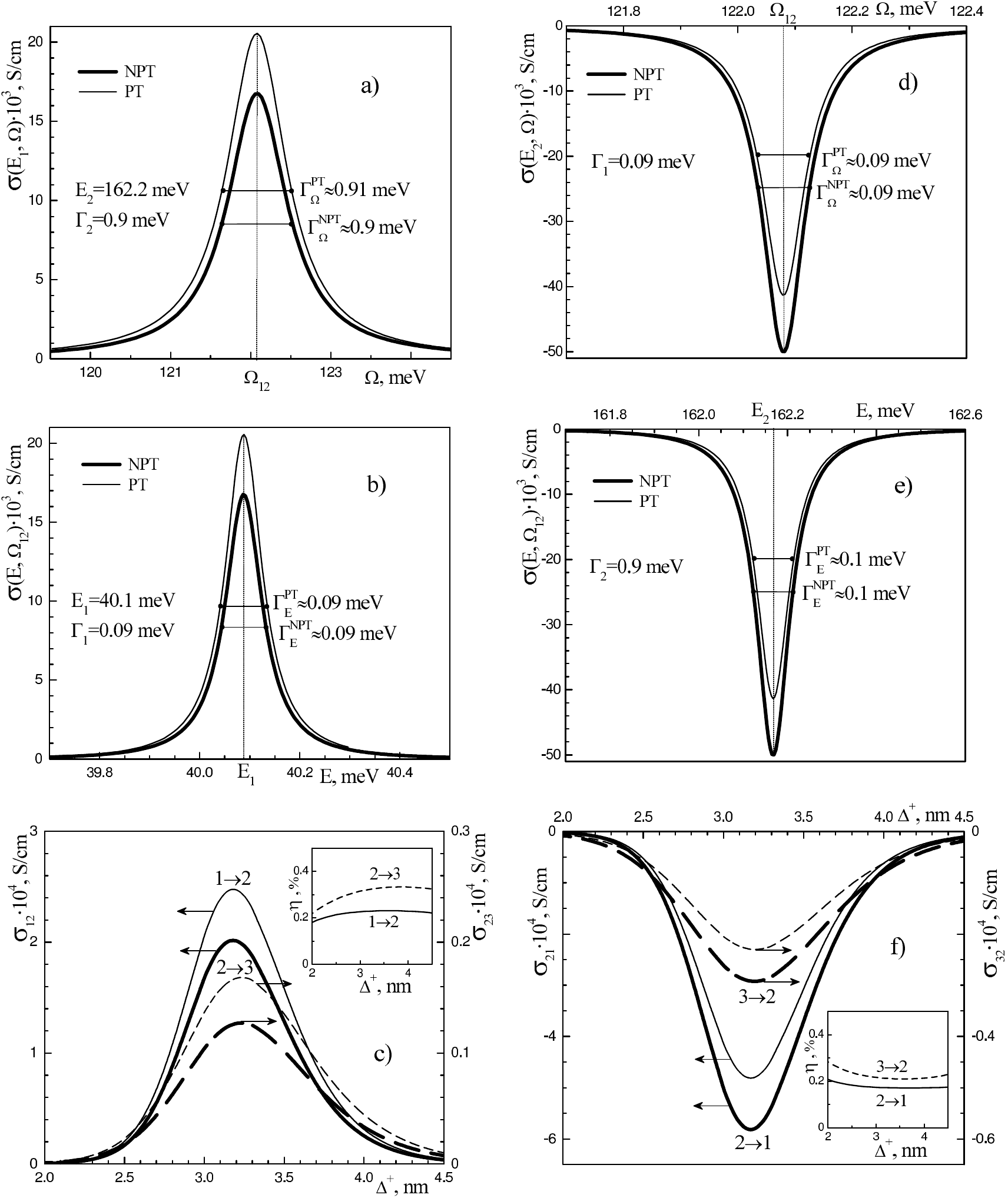}}
\caption{Dependences of maximal magnitudes of positive (a, b, c)
and negative (d, e, f) conductivities $\sigma$ on electromagnetic
field energy $\Omega = \hbar \omega$ (a, d) and electron energy
$E$ (b, e) in two-barrier RTS at $\Delta^{+}=\Delta^{-}$=3 nm.
Dependences of maximal magnitudes of positive and negative
conductivities on the relationship between the both barrier widths
(c, f) at $\Delta^{+}+\Delta^{-}$= 6~nm obtained within the
non-perturbation theory (bold solid and dashed curves) and
perturbation theory (thin solid and dashed curves).}
\label{fig2}
\end{figure}

\section{Conclusions}

1. The non-perturbation theory of active dynamic conductivity for the
open two-barrier RTS, preserving the terms linear over the electromagnetic
field in an electron wave function, is proposed for the first time.

2.  It is shown that the properties of positive and negative
conductivities of two-barrier RTS, shown earlier within the
linear approximation over the field perturbation theory, in the so-called weak
signal approximation are not only qualitatively similar but
quantitatively correlate to the results of a more exact
non-perturbation theory proposed.

3.  The developed non-perturbation theory of dynamic conductivity
for the two-barrier nano-structure can be used for the multi-shell
RTS and generalized for the physically and technically important case
of electron currents interacting with strong electromagnetic fields
in quantum cascade lasers and detectors.

\newpage

\ukrainianpart

\title{Непертурбаційна теорія електронної динамічної провідності двобар'єрної резонансно-тунельної наноструктури}
\author{М.В.Ткач, Ю.О.Сеті, О.М.Войцехівська}
\address{Чернівецький національний університет ім. Ю.Федьковича, вул. Коцюбинського, 2, 58012 Чернівці, Україна}

\makeukrtitle

\begin{abstract}
\tolerance=3000%
Вперше запропоновано непертурбаційну теорію електронної динамічної
провідності відкритої двобар'єрної резонансно-тунельної структури
у моделі прямокутних потенціалів і різних ефективних мас
електронів у різних елементах наносистеми та з лінійною за
напруженістю електромагнітного поля хвильовою функцією системи.
Показано, що результати розвинутої раніше теорії динамічної
провідності у малосигнальному наближенні (у межах
теорії збурень) якісно і кількісно корелюють з отриманими
результатами. Переваги непертурбаційної теорії в тому, що вона
може бути поширена на випадок взаємодії потоків електронів з
потужними електромагнітними полями у відкритих багатошарових
резонансно-тунельних наноструктурах, як активних елементах
квантових каскадних лазерів і детекторів.
\keywords резонансно-тунельна наноструктура, провідність, непертурбаційна теорія

\end{abstract}

\end{document}